\documentclass[12pt,a4paper]{article}

\usepackage{amsmath}
\usepackage{amssymb}
\usepackage{mathrsfs}
\usepackage{graphicx,subcaption}
\usepackage{dsfont}
\usepackage[pdftex,colorlinks=true,linkcolor=black,citecolor=black,urlcolor=black]{hyperref}
\usepackage{enumerate}
\usepackage{epstopdf}
\usepackage{slashed}
 \usepackage[normalem]{ulem}

\usepackage[greek,english]{babel}

\numberwithin{equation}{section}

\def\eq#1 { \begin{equation} #1 \end{equation} }
\def\eqn#1{ \begin{align} #1 \end{align} }

\def\Re{{\rm Re}\,}

\def\sl2r{SL(2,\mathbb{R})}

\newcommand{\lsim}{\mathrel{\hbox{\rlap{\lower.55ex \hbox{$\sim$}} \kern-.3em \raise.4ex \hbox{$<$}}}}
\newcommand{\gsim}{\mathrel{\hbox{\rlap{\lower.55ex \hbox{$\sim$}} \kern-.3em \raise.4ex \hbox{$>$}}}}

 \newcommand{\be}{\begin{equation}}
\newcommand{\ee}{\end{equation}}

\newcommand{\PT}{$\mc{PT}$}

\usepackage{bm}

\usepackage{mathtools,cancel}
\usepackage[usenames,dvipsnames,svgnames,table]{xcolor}

\usepackage{amsmath}
\newmuskip\pFqmuskip

\newcommand*\pFq[6][8]{%
  \begingroup 
  \pFqmuskip=#1mu\relax
  \mathcode`\,=\string"8000
  \begingroup\lccode`\~=`\,
  \lowercase{\endgroup\let~}\pFqcomma
  {}_{#2}F_{#3}{\left[\genfrac..{0pt}{}{#4}{#5};#6\right]}%
  \endgroup
}
\newcommand{\pFqcomma}{\mskip\pFqmuskip}

\newmuskip\pPqmuskip

\newcommand*\pPq[6][8]{%
  \begingroup 
  \pPqmuskip=#1mu\relax
  \mathcode`\,=\string"8000
  \begingroup\lccode`\~=`\,
  \lowercase{\endgroup\let~}\pPqcomma
  {}_{#2}\Psi^*_{#3}{\left[\genfrac..{0pt}{}{#4}{#5};#6\right]}%
  \endgroup
}
\newcommand{\pPqcomma}{\mskip\pPqmuskip}

\DeclareMathOperator{\Tr}{Tr}

\newcommand{\sign}{\text{sign}}
\newcommand{\mbb}{\mathbb}
\newcommand{\mc}{\mathcal}

\renewcommand{\Re}{\text{Re}}

\newcommand{\penguin}{\includegraphics[width=0.2in]{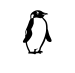}}

\usepackage{marvosym}
\usepackage{tikz}
\usepackage{tikzsymbols}
\usepackage{phaistos}
\makeatletter
\def\@fnsymbol#1{\ensuremath{\ifcase#1\or \,\or \text{\Cat}  \or
   \mathsection\or \mathparagraph\or \|\or **\or \dagger\dagger
   \or \ddagger\ddagger \else\@ctrerr\fi}}
\makeatother
\renewcommand{\thefootnote}{\fnsymbol{footnote}}
\begin{document}

\title{\begin{flushright}\vspace{-1in}
       \mbox{\normalsize  Imperial/TP/2018/LA/01}
       \end{flushright}
       \vskip 20pt
Mass deformed ABJM and $\mc{PT}$ symmetry
}

\date{\today}

\author{
     Louise Anderson\penguin  \footnote{ \hspace{-4mm}\protect \penguin \hspace{-1mm}\href{mailto:louise.m.a.anderson@imperial.ac.uk}
 { louise.m.a.anderson@imperial.ac.uk}}~
 \\
  Matthew M. Roberts
   \footnote{\href{mailto:matthew.roberts@imperial.ac.uk}
     {matthew.roberts@imperial.ac.uk}}      \\ 
   {\it \it Blackett Laboratory, Imperial College}\\
   {\it   London, SW7 2AZ, U.K. }
}

\maketitle

\begin{abstract}
We  consider real mass and FI deformations of ABJM theory preserving supersymmetry in the large $N$ limit, and compare with holographic results. On the field theory side, the problems amounts to a spectral problem of a non-Hermitian Hamiltonian. For certain values of the deformation parameters this is invariant under an antiunitary operator (generalised $\mc{PT}$ symmetry), which ensures the partition function remains real and allows us to calculate the free energy using tools from statistical physics. The results obtained are compatible with previous work,  the important new feature being that  these are obtained directly from the real deformations, without analytic continuation.
\end{abstract}

\tableofcontents

\renewcommand{\thefootnote}{\arabic{footnote}}
\setcounter{footnote}{0}


\section{Introduction}\label{sec:intro}

The low energy dynamics on a stack of $N$ coincident $M_2$ branes on a $\mbb{Z}_k$ orbifold is conjectured to be described by the ABJM theory \cite{Aharony:2008ug}, a Chern-Simons field theory, whose gravitational dual is 11D supergravity on backgrounds which are asymptotically $AdS_4\times S^7/\mbb{Z}_k$. This duality has undergone a number of tests, including the derivation of the famous $N^{3/2}$ scaling of the degrees of freedom of $N$ coincident $M_2$ branes \cite{Klebanov:1996un} from the field theory side \cite{Drukker:2010nc}. 

ABJM theory can be deformed by relevant operators, including real masses for the bifundamental scalars and Fayet-Iliopoulos parameters \cite{Hosomichi:2008jd, Hosomichi:2008jb, Gomis:2008vc}. On the gravitational side, the deformation corresponds to modifying boundary conditions.

In this paper we calculate the Euclidean free energy of ABJM with real mass and FI deformations both directly in the field theory and via its gravitational dual. The breaking of conformal symmetry leads to a partition function which is a nontrivial function of deformation parameters.
Using supersymmetric localisation \cite{Pestun:2007rz}, the path integral for three-dimensional theories with $\mathcal{N}\geq 2$ supersymmetry on a three-sphere can be computed exactly  \cite{Kapustin:2009kz, Jafferis:2010un, Hama:2010av}, resulting in a matrix model. This has enabled a large amount of progress to be made, see for instance \cite{Pestun:2016zxk} for a review and extensive references.  In particular, it allows us to write the three-sphere partition function of the deformed ABJM theory as a matrix model \cite{Drukker:2010nc,  Kapustin:2010xq}.

It was shown in \cite{Marino:2011eh} that the matrix model for the undeformed ABJM theory can be recast as a partition function for a one dimensional gas of $N$ noninteracting fermions with a complicated Hamiltonian, reducing it to a problem in statistical physics. 
When adding real masses to the theory, this Hamiltonian becomes non-Hermitian,  creating complications in the standard techniques for these quantum mechanical systems. However, for certain values of the deformation parameters, the system is invariant under a combined parity and time reversal transformation, called $\mc{PT}$ symmetry \cite{Bender:1998ke, Bender:2002yp}.\footnote{In some cases this is generalised to a combination of \PT~and a unitary transformation, called generalised \PT~symmetry.} In such theories, the eigenvalues of the Hamiltonian are either real (\PT~invariant states) or come in complex conjugate pairs (states which map to each other under \PT), and the partition function remains real in both phases.
For general Chern-Simons theories, the three-sphere free energy is a priori complex, but here \PT~symmetry forces the imaginary part to vanish. For more general deformation parameters, while the Hamiltonian is not \PT~invariant, the spectral $Z$ functions derived from it are, which again guarantees a real partition function.\footnote{For more general quiver Chern-Simons theories, there exist real mass deformations which lead to Hermitian spectral problems \cite{Drukker:2015awa}, but this is not possible in ABJM.}

To compare with gravitational results, we are interested in the partition function at large $N$ and fixed $k$,\footnote{For a review of localisation in large $N$  3d Chern-Simons theories, see \cite{Marino:2016new} and references therein. } corresponding to the classical limit of the quantum  system.
At leading order in $N$, we find a perfect agreement between the holographic and the field theory calculations, which are related to the results of \cite{Freedman:2013ryh,Nosaka:2015iiw} via analytic continuation. In the field theory, this analytic continuation is subtle, and it requires careful tracking of the branch cuts of the logarithm over the complex plane during the calculations. Our derivation is valid for all real values of the mass deformations.

The rest of this paper is outlined as follows:
In section \ref{sec:abjm}, we review ABJM theory and the mass deformations we consider. In section \ref{sec:grav} we review the holographic calculation of $F$ maximisation and relate it, via analytic continuation, to a real mass deformation. In section \ref{sec:Fermi}, using  results of the now standard  localisation calculations for three-dimensional $\mc{N}\geq 2$ theories, we rewrite the partition function as a gas of $N$ noninteracting fermions, and find the corresponding Hamiltonian as a function of the deformation parameters. We then outline the standard technique to obtain the large $N$ behaviour of the free energy.  As mentioned, our Hamiltonian is however \emph{not} Hermitian for non-vanishing deformation parameters, but in some cases it is $\mc{PT}$ symmetric, and we review the implication of this in section \ref{sec:PT}.  We are then finally set to compute the three-sphere free energy in the large $N$ limit in section \ref{sec:FreeEnergy}. We conclude with a discussion and open questions in section \ref{sec:Discussion}.

As this letter finished preparation the following appeared \cite{Honda:2018cky}, which has some overlap with our analysis. Our calculation however requires no saddlepoint approximation and we see no sign of supersymmetry breaking on either side of the duality. This discrepancy should be investigated further, particularly in the nonperturbative contributions.

\section{ABJM theory and real mass deformations}\label{sec:abjm}
ABJM theory is a Chern-Simons theory with gauge group $U(N)_{k}\times U(N)_{-k}$. This theory has $\mc{N}=6$ supersymmetry \cite{Aharony:2008ug}, though for $k=1,2$, this is enhanced to $\mc{N}=8$ \cite{Benna:2009xd}. It will be convenient for us to work in  $\mc{N}=2$ superspace formalism.

The ABJM theory contains one $\mc{N}=4$ vector multiplet transforming in the adjoint  of the respective $U(N)$ gauge groups 
and two hypermultiplets in the bifundamental.  The components of the multiplets  are presented in table \ref{tab:fieldcont} below.  (here $\chi, \lambda$ are two-component Dirac spinors, $\sigma$ is a real scalar and $\varphi$ a complex scalar).
 
 \begin {table}[h!]
\begin{center}
\renewcommand{\arraystretch}{1.5} 
\begin{tabular}{|r| l| l|  }
\hline
 $\mc{N}=4$ & $\mc{N}=2$   & Components  
 \\
  \hline    \hline
Vector &  Vector  & $\mc{V}=(A_\mu, \lambda, \sigma, D)$  
 \\
 \cline{2-3}
 & Chiral  & $\Phi = (\varphi, \chi, F_{\Phi})$ 
 \\ 
  \hline
    \hline
Hyper &  Chiral  & $Z=(Z,  \Psi, F)$ 
  \\
 \cline{2-3}
 & Chiral  & $W=(W, \tilde{ \Psi}, \tilde{F})$ 
   \\ 
  \hline
\end{tabular}
  \caption{The field content of $\mc{N}=4$ and $\mc{N}=2$ multiplets in three dimensions.}
   \label{tab:fieldcont}
\end{center}
\end{table}

The Lagrangian of ABJM on $S^3$ (with unit radius) is given by a supersymmetric Chern-Simons kinetic  terms for each of the $\mc{N}=2$ vector multiplets with levels $k$ and $-k$,
\begin{equation*}
S_{CS} =
 \int d^3 x 
\Tr \left(
 A\wedge dA
 +
 \frac{2i}{3} A\wedge A\wedge A 
 - \bar{\lambda}\lambda+2D\sigma
 \right)
,\end{equation*}
where the trace is assumed to be appropriately normalised with the levels to be invariant under large gauge transformations and standard terms are included for the fermions. The kinetic terms and interactions for the bifundamental chiral multiplets are  given by: 
 \begin{equation}
 \label{eq:chiral_action}
S_{chiral}=\int d^3x \sqrt{g}\Big(
\,
D_\mu \bar{ Z} D^\mu  Z
-i \bar{  \Psi} \gamma^\mu D_\mu   \Psi
+ \frac{3}{4 } \bar{ Z}  Z
- i \bar{  \Psi} \sigma   \Psi
+ i \bar{  \Psi} \lambda  Z
- i \bar{ Z} \bar{\lambda}   \Psi
+ i \bar{ Z} D  Z
+\bar{ Z} \sigma^2  Z
+ \bar{F}F
\Big).
\end{equation}
The famous quartic superpotential is needed to ensure supersymmetry enhancement from $\mc{N}=2$ to $\mc{N}=6$.

The real mass deformations can be thought of as coupling to background vector multiplets with a supersymmetric expectation value.  Such an expectation value can be parametrised by one parameter, which, if real, corresponds to giving real masses to the hypermultiplet scalars via the quadratic coupling in \eqref{eq:chiral_action}.\footnote{If this is instead taken to be purely imaginary, the resulting terms in the Lagrangian give a shift to nonstandard R-charge assignments. The partition function is conjectured to be an analytic function of this  parameter \cite{Closset:2012vg}.} It is also possible to introduce Fayet-Iliopoulos terms, modifying the Lagrangian by:
\begin{equation*}
\mc{L}_{FI} = \frac{k}{2\pi} \zeta \, \Tr(D^{(1)}+D^{(2)} -\left( \sigma^{(1)}+\sigma^{(2)}\right))
,\end{equation*} 
where  $D^{(1,2)}$ are the auxiliary scalars in the vector multiplets in the two nodes.
After integrating out the auxiliary scalar in the vector multiplet acts as a shift in the expectation value of the vector multiplet scalars via
\begin{align*}
\sigma^{(1)} \rightarrow \sigma^{(1)}+ \frac{\zeta}{2}
, \qquad \sigma^{(2)} \rightarrow \sigma^{(2)}-\frac{\zeta}{2}
.\end{align*}

By a change of integration variables in the localisation calculation, one can see that these deformations in the most general case can be thought of as giving two different masses, $m_{1,2} = m \pm 2 \zeta$ to the dynamical scalars of the theory.

These two parameters incorporates all possible real mass deformations of the theory which are accessible on $S^3$ via localisation calculations. There is a third mass deformation which breaks supersymmetry to $\mc{N}=1$.



\section{Holographic results }\label{sec:grav}

The R-charges  of the scalars in the $\mc{N}=2$ chiral multiplets are given in terms of three parameters $\delta_i$ in \cite{Freedman:2013ryh}, and are related to the mass deformations via:
\begin{align}\label{eq:r_charge_shifts}
R[Z^1]=&\frac{1}{2}+\delta_1+\delta_2+\delta_3 = i m_{Z^1}+\frac{1}{2} \nonumber
\\
R[W_1]=&\frac{1}{2}-\delta_1+\delta_2-\delta_3 = i m_{W_1}+\frac{1}{2} 
\\
R[Z^2]=&\frac{1}{2}+\delta_1-\delta_2-\delta_3 = i m_{Z^2}+\frac{1}{2} \nonumber
\\
R[W_2]=&\frac{1}{2}-\delta_1-\delta_2+\delta_3 = i  m_{W_2}+\frac{1}{2}\nonumber
\end{align}
Note that marginality of the superpotential constrains these deformations to only be a function of the three parameters $\delta_{i}$, two of which are accessible through the localisation calculation. In our case, the most general deformation corresponds to half of the scalars having mass $\pm m_1$ and the other half $\pm m_2$:
\begin{align*}
m_{Z^1}=& m_1 \qquad,\qquad
m_{W_1}= -m_1
\\
m_{Z^2}=& m_2 \qquad,\qquad
m_{W_2}= -m_2
\end{align*}
leading to, as expected, coupling the background theory to two background vector multiplet where the expectation values of the background scalars are given by $i\delta_a$ such that:
{\small
\begin{align*}
\delta_1 = \frac{i}{2} \left(m_1+m_2\right) = i m
\quad,\quad
\delta_2=0
\quad,\quad
\delta_3= \frac{i}{2} \left(m_1-m_2\right) =2 i \zeta .
\end{align*}
}
In \cite{Freedman:2013ryh} the free energy of the theory deformed by \eqref{eq:r_charge_shifts} was computed holographically for general $\delta_a$. The free energy is conveniently expressed in terms of the parameters $c_a$ as
\begin{align*}
F=\frac{  \pi  L^2 }{ 2 G_4 }
\frac{ \left(1-c_1^2\right) \left(1-c_2^2\right) \left(1-c_3^2\right) }{\left(c_1 c_2 c_3+1\right){}^2}
\end{align*}
related to the deformation parameters $\delta_a$ via
\begin{align*}
\delta _a=\frac{\frac{c_3 c_2 c_1}{c_a}+c_a}{2 \left(1+c_1 c_2 c_3\right)}
.\end{align*}

Solving these gives a quadratic expression for the $c_a$'s in terms of the $\delta_a$'s, which, when the relation to the real mass deformations is used leads to the following expression for the free energy:
\begin{align}
F=\frac{\pi  L^2 }{2 G_4}\sqrt{\left(1+[2 m_1]^2\right) \left(1+[2 m_2]^2\right)}.
\end{align}
While the calculation in \cite{Freedman:2013ryh} was done in the $\mathcal{N}=8$ supergravity theory dual to $k=1$, it involved only the metric and scalars neutral under the $SO(8)$ R-symmetry. The $\mathbb{Z}_k$ quotient then acts trivially on the 11D solution. We can therefore generalise to arbitrary $k$ by noting that the quotient simply rescales $L$ and $G_4$, giving  $\frac{\pi  L^2 }{2 G_4}=\frac{\pi  \,\sqrt{2  k}   \, N^{3/2}}{3}$. We will find that this 
agrees perfectly with the field theory result in \eqref{eq:F_loc_general} at leading order in $N$.



\section{The ABJM matrix model as a Fermi gas\label{sec:Fermi}}

Using  standard techniques of supersymmetric localisation  \cite{Pestun:2007rz}, one can show that the partition function of $\mathcal{N}\geq 2$ theories on $S^3$ localises onto constant field configurations for the vector multiplet scalars, $\sigma$. 
Using the rules presented in for example \cite{Kapustin:2009kz, Drukker:2010nc, Kapustin:2010xq, Gulotta:2012yd}, we can see that each $U(N)$ vector multiplet, with Chern-Simons level $k$ and FI-parameter $\zeta$, contributes to the partition function with
\begin{equation*}
\int
  d^N\hspace{-1mm}\sigma 
\prod _{i<j}^N \left[2\sinh\left(\pi\left(\sigma _i-\sigma _j\right)\right) \right]^2 e^{\pi i k \sum _{i=1}^N  \left[ \sigma _i^2 -2 \zeta  \sigma _i \right]}
\end{equation*}
while each $\mc{N}=4$ hypermultiplet with mass $m$ transforming as $\left(\mathbf{r} , \bar{\mathbf{r}} \right)$ of $U(N)_{k}\times U(N)_{-k}$ results in a 1-loop contribution to the partition function 
\begin{equation*}
\frac{1}{\prod _{i,j}^N 
2\cosh \left( \pi \left(\sigma^{(1)} _i-\sigma^{(2)} _j  -m  \right)\right)}
,\end{equation*}
where $\sigma^{(1,2)}$ are the vector multiplets in the two nodes. Furthermore, there is an overall normalisation factor of $\frac{1}{N!}$ for each $U(N)$ factor of the gauge group.

We can therefore easily obtain the partition function for the deformed ABJM theory described in section \ref{sec:abjm} as:
{\footnotesize
\eqn{
\label{eq:MM_org}
Z_{ABJM}= \frac{1}{2^{2N}(N!)^2  }
\int
  d^N\hspace{-1mm}\sigma^{(1)}  \,d^N \hspace{-1mm} \sigma^{(2)}& \,
  {\small
\frac{\prod _{i<j}^N\sinh^2 \left(\pi\left(\sigma^{(1)} _i-\sigma^{(1)} _j\right)\right) \sinh^2 \left( \pi\left(\sigma^{(2)} _i-\sigma^{(2)} _j\right)\right) }{\prod _{i,j}^N 
\cosh \left( \pi \left(\sigma^{(1)} _i-\sigma^{(2)} _j  -m  \right)\right)
\cosh \left( \pi\left(\sigma^{(1)} _i-\sigma^{(2)} _j  +  m  \right)\right)
}
}\nonumber \\
&\times e^{\pi i k \sum _{i=1}^N  \left[ \left(\sigma^{(1)} _i\right)^2-\left(\sigma^{(2)} _i\right)^2 -2 \zeta  \left(\sigma^{(1)} _i+\sigma^{(2)} _i\right) \right]}
.}
}
Using a change of integration variables, $\mu_i=\sigma^{(1)}_i-\zeta$, $\nu_i=\sigma^{(2)}_i+\zeta$, it is easy to see that a deformation with both mass $m$ for the hypermultiplets as well as FI term $\zeta$ for the vector multiplet is equivalent to deforming the theory by giving two different masses to the hypermultiplets: half of them are given mass $m_1$ and the other half $m_2$, defined by
\begin{equation}
\label{eq:def_param}
 m_1=  m+ 2 \zeta  
\qquad,\qquad
 m_2 =m- 2 \zeta  
\end{equation}
and the partition function can be written as:
{\footnotesize
\begin{equation}
\label{eq:MM_diff_masses}
Z_{ABJM}= \frac{1}{2^{2N}(N!)^2  }
\int
  d^N\hspace{-1mm}\mu \,d^N \hspace{-1mm} \nu\,
  {\small
\frac{\prod _{i<j}^N \sinh^2 \left(\pi\left(\mu_i-\mu_j\right)\right)  \, \sinh^2 \left( \pi\left(\nu_i-\nu_j\right)\right) }{\prod _{i,j}^N 
\cosh \left( \pi \left(\mu_i-\nu_j  +m_1  \right)\right)
\cosh \left( \pi\left(\mu_i-\nu_j  -  m_2  \right)\right)
}
}
e^{\pi i k \sum _{i=1}^N  \left[ \mu_i^2-\nu_i^2  \right]}
.\end{equation}
}



To make contact with the holographic calculations in section \ref{sec:grav}, we want to solve the matrix model of \eqref{eq:MM_org} at large $N$ for fixed value of $k$, sometimes known as the M-theory limit.\footnote{For investigations of the mass deformed theory in the 't Hooft limit, see \cite{Anderson:2014hxa, Anderson:2015ioa, Nosaka:2015bhf, Nosaka:2016vqf,Honda:2018cky}}   This limit is conveniently accessible by following \cite{Marino:2011eh} to recast the matrix model in \eqref{eq:MM_org}  in terms of a gas of $N$ non-interacting fermions with a complicated Hamiltonian. The main step for this is to use Cauchy's determinant identity
\begin{equation*}
\frac{\prod _{i<j}^N \sinh \left(\mu _i-\mu _j\right) \sinh \left(\nu _i-\nu _j\right)}{\prod _{i,j}^N \cosh \left(\mu _i-\nu _j\right)}
=
\sum_{\sigma\in S_N}(-1)^\sigma
\,
\frac{1}{\prod _i \cosh \left(\mu _i-\nu _{\sigma (i)}\right)}
\end{equation*}
to rewrite \eqref{eq:MM_org} as 
\begin{align}
Z=
 \frac{1}{2^{2N}N! \, }
\sum_{\sigma} (-1)^{\sigma}
\int
  d^N\mu \, d^N  \nu \;
\frac{
e^{\pi i k \sum _{i=1}^N \left(\mu _i^2-\nu _i^2\right)}
}{
\prod_{i}
\cosh\pi \left(\mu _i-\nu _i  +  m_1   \right)
\cosh\pi \left(\mu _i-\nu _{\sigma (i)}  -  m_2  \right)
}
,\end{align}
where  we have used that the integral only depends on the composition $\sigma \circ \sigma'$. The sum over permutations with alternating signs here is the key ingredient to reinterpreting the problem as a gas of fermions.
Using now that $\frac{1}{\cosh}$ is it's own Fourier transform,
\begin{equation*}
\frac{1}{\cosh \left( \pi \left(\mu _i-\nu _i\right)\right)}
=
\int d \tau _i
\frac{e^{ 2 \pi i \tau _i \left(\mu _i-\nu _i\right)}}{\cosh \left( \pi \tau _i\right)}
,\end{equation*}
 we can simplify the expressions even further.  This rewriting turns the integrals over $\mu$ and $\nu$ into Gaussians which can be carried out straight-forwardly, resulting in
 {\footnotesize
 \begin{align}
Z_{ABJM}=&  \,
 \frac{1}{ N!  }
\sum_{\sigma} (-1)^{\sigma}
\int
 d^N  \tau\;
 \prod_{i} 
 \frac{e^{  -\pi i  m_2  \tau_i   }
}{
\left(2\cosh \left( \pi \tau _i\right)\,\right)^{1/2}
}
\,
  \frac{1}{
2k\cosh\Big( \frac{\pi}{k}
\left( \tau_i- \tau_{\sigma(i)} -k m_1  \right)
 \Big)
 }
 \frac{e^{  -\pi i m_2     \tau_{\sigma(i)}} 
}{
\left(2 \cosh \left( \pi \tau_{\sigma(i) }\right) \, \right)^{1/2}
}
\label{eq:part_func1}
,\end{align}
}
where in the last step we have used $(-1)^{\sigma(i)}=(-1)^{\sigma^{-1}(i)}$ and the inverse Fourier transform.
The expression \eqref{eq:part_func1} allows us to interpret the system as an ordinary quantum mechanical system: that of a gas of $N$ free fermions described by a one-particle density matrix $\rho$, which in position space reads:
\begin{align}
\nonumber
\rho(x_1,x_2)=&\,
  \frac{e^{  -i \pi   m_2  x_1  }
}{
\left(2 \cosh \left( \pi  x_1 \right)\,\right)^{1/2}
}
\,
  \frac{1}{
 2k \cosh\Big( \frac{\pi}{k}
\left( x_1- x_2 -k m_1 \right)
 \Big)
 }
 \frac{e^{- i \pi   m_2     x_2}
}{
\left(2\cosh \left( \pi x_2 \right)\,\right)^{1/2}
}  =&\,
\langle x_1|\hat{\rho} |x_2 \rangle .
  \label{eq:density}
\end{align}
We can as usual choose to describe the system in terms of a Hamiltonian $H$,\footnote{We here use eigenstates with the canonical normalisation $\langle x|x'\rangle=\delta(x-x')$, $\langle p|p'\rangle=\delta(p-p')$ and $\langle x | p \rangle = \frac{1}{\sqrt{2 \pi \hbar}}e^{i p x /\hbar }$}
\begin{equation}
\label{eq:Hamiltonian_def}
\hat{\rho} = e^{-\frac{1}{2}U(x)} e^{-T(p)} e^{-\frac{1}{2}U(x)} 
   \;  =\,  e^{-\hat{H}}
\end{equation}
for operators 
\begin{align*}
U(x)
= 
 \log \left[2  \cosh \pi x  \right]
+2 \pi i \,  m_2 x
\qquad,\qquad
T(p)  =
\log\left[2
\cosh \pi p
\right]
+ 2 \pi i \,  m_1 p
,\end{align*}
and $x, p$ canonical conjugate variables with canonical commutation relations where the Chern-Simons level plays the role of Planck's constant:
\begin{align*}
[x,p]=i \hbar 
\qquad,\qquad \hbar= \frac{k}{2 \pi}
.\end{align*}
 In terms of the density matrix, the partition function takes the remarkably simple form:
\begin{align}
Z=
 \frac{1}{N!  }
\sum_{\sigma} (-1)^{\sigma}
\int
 d^N z\;
    \prod_{i}
 \rho(z_i,z_{\sigma(i)})
.\end{align}


\subsection{The semiclassical limit: WKB expansion \label{sec:WKB}}

The Hamiltonian defined via \eqref{eq:Hamiltonian_def} is very complicated, and it's energy levels are not known.
 In particular, it is not Hermitian, something we will comment on more in the next section. Let us for now ignore this issue, and review the approach ordinarily taken in this situation: doing a WKB approximation about $\hbar=0$. Conveniently enough, large $N$ corresponds to the semiclassical limit, so the expansion in $\hbar$ corresponds to doing a perturbative expansion in $\frac{1}{N}$ \cite{Marino:2011eh}. It is convenient to use Wigner's phase space formalism \cite{Wigner:1932aa}, and consider the Wigner transform\footnote{For a review of phase space approach to quantisation, see for example \cite{1984PhR...106..121H}.
} of the density operator, and then using Baker-Campbell-Hausdorff on the exponential in \eqref{eq:Hamiltonian_def} to obtain the Hamiltonian. In \cite{Marino:2011eh}, it was shown that the Wigner-transform of the Hamiltonian takes the form
\begin{align}
\label{eq:Hw}
H_{W}(x,p)=H_{cl.}-\frac{\hbar^2}{12} \,T'(p)^2U''(x) +\frac{\hbar^2}{24} \,U'(x)^2T''(p) +\mc{O}(\hbar^4)
\end{align}
where $H_{cl.}$ is the classical Hamiltonian:
\begin{align}
\label{eq:Hcl}
H_{cl.}=& T(p)+U(x)  
.\end{align}
At large $N$, the leading term will be fully determined by the classical contribution to $H_W$.

It will be convenient to work in the grand canonical ensemble, where the partition function is obtained from the grand canonical potential, $J(\mu)$ by the inverse transform:
\begin{equation}
\label{eq:ZinJ}
Z(N)=\frac{1}{2\pi i} \int d\mu\,  e^{J(\mu)-\mu N}
\end{equation}
and
\begin{align}
\label{eq:Grand_can_pot}
J(\mu) = - \sum_{l\geq 1} Z_l \frac{(-z)^l}{l}
\qquad ,\qquad
Z_l=\Tr e^{-l H}\;  
.\end{align}
The $Z_l$ are often referred to as spectral $Z$ functions. We identify $z=e^\mu$ as a fugacity with corresponding chemical potential $\mu$. This sum is only convergent for small $|z|$, but can under favourable conditions be analytically continued to the entire complex plane. 

The small $|z|$-limit here corresponds to small $N$, but as we wish to make connections to holography, we are interested in the large $N$ limit. In this case, to avoid the analytic continuation necessary in \eqref{eq:Grand_can_pot}, it is more convenient to consider the Mellin-Barnes representation of the grand canonical potential  $J(\mu)$ \cite{Hatsuda:2015oaa}:
\begin{align}
\label{eq:J_MBrep1}
J(\mu)=-\frac{1}{2\pi i} \int_{c-i\infty}^{c+i\infty}
d l\, \Gamma(l) \Gamma(-l) Z_l \, e^{l \mu}
.\end{align}
Here, $c$ is a constant that must lie between zero and the (real part of the) first pole of the integrand in the right half-plane. Provided that the spectral $Z$ functions have no poles in the right half plane, we have $0<c<1$. This is not a priori true, and \eqref{eq:J_MBrep1} is therefore valid only if the spectral $Z$ functions have the appropriate pole structure. For  $\mu<0$, one can close the integration contour in the right half plane, and the residue formula then recovers \eqref{eq:Grand_can_pot}, whereas for $\mu>0$, however, one can close the contour in the left half-plane instead, obtaining 
\begin{align}
\label{eq:J_MBrep}
J(\mu)=-\sum_{\underset{\Re(l)<c}{\text{poles with}}}
\text{Res} \left\{\, \Gamma(l) \Gamma(-l) Z_l \, e^{l \mu}\right\}
.\end{align}
As $\mu\rightarrow \infty$, all other poles than at $s=0$ are exponentially suppressed in $\mu$, and this will give us the leading behaviour in $N$.

The Wigner phase space formalism allows us to, in principle, to go beyond the semiclassical limit and compute the partition function to all orders in $\hbar$. The quantum corrections to the Hamiltonian \eqref{eq:Hcl} arising from the fact that $x,p$ don't commute results in a series expansion of the spectral $Z$ functions in $\hbar$ as
\begin{equation}
\label{eq:Zl_series}
Z_l=\frac{1}{\hbar}\sum_{n=0}^{\infty}Z_{l}^{(n)}\hbar^{2n}
.\end{equation}
 In \cite{Marino:2011eh}, in the undeformed case it was shown that only the first $\hbar^2$-corrections contribute to the asymptotic series in $\frac{1}{N}$ to the free energy.\footnote{For the undeformed theory, all quantum corrections to the grand canonical potential, including non-perturbative effects, were the result of a combined effort of \cite{Marino:2011eh, Hatsuda:2012hm, Putrov:2012zi, Hatsuda:2012dt, Calvo:2012du, Hatsuda:2013gj} and eventually presented in  \cite{Hatsuda:2013oxa}.} In appendix \ref{sec:quantum}, we  show that the leading quantum correction is independent of the deformation parameters, and we will therefore ignore them for now and only consider the semiclassical limit.

Recall that the strict $N \rightarrow \infty$ limit corresponds to the classical limit of the quantum system, and we can use tools from standard classical mechanics. In particular, it will be useful for us to notice that in this limit, the  spectral $Z$ functions $Z_l$ are completely determined by $Z^{(0)}$, and can be computed as an integral over real phase space:
\begin{equation}
\label{eq:Zl}
N \rightarrow \infty ,~ Z_l  = \; 
\frac{1}{2 \pi \hbar }\int_{-\infty}^{\infty} \hspace*{-2mm}dp \,\int_{-\infty}^{\infty} \hspace*{-2mm} 
dx \, e^{-l H_{cl.} (x,p)}
,\end{equation}
with the only subtlety that this expression is valid for \emph{Hermitian} $H_{cl.}$. Overcoming this difficulty  will be the topic of the next section.

\section{$\mc{PT}$ symmetric quantum and statistical mechanics }\label{sec:PT}
We will now tackle the issue of Hermiticity. For nonzero deformation parameters  $ m_1$ and $ m_2$ (which are related to mass and FI parameters via \eqref{eq:def_param}), the Hamiltonian defined via \eqref{eq:Hamiltonian_def}  is not Hermitian. Hermitian versions of our systems have been studied previously in the literature \cite{Nosaka:2015iiw, Drukker:2015awa}, but the analytic continuations are subtle at best, and it is interesting to study the complex Hamiltonian  \eqref{eq:Hamiltonian_def} directly.   Normally, the Hermiticity of the density matrix (or Hamiltonian) is used to guarantee real eigenvalues, but  this is not a necessary condition. There are also families of non-Hermitian Hamiltonians with real eigenvalues, perhaps the most famous one being the deformation of the harmonic oscillator,  $p^2+x^2(ix)^\epsilon$ ($\epsilon\geq0$). These Hamiltonians are  invariant under the combination of parity and time reversal, and as such said to be $\mc{PT}$ symmetric \cite{Bender:1998ke}. The parity $\mc{P}$  and time reversal $\mc{T}$ operators act on $\hat{x},\hat{p}$ as:
\begin{align*}
\mc{P} \hat{x} \mc{P} = -\hat{x}
\qquad
\mc{P} \hat{p} \mc{P} = -\hat{p}
\quad,\quad
\mc{T} \hat{x} \mc{T} = \hat{x}
\qquad
\mc{T} \hat{p} \mc{T} = -\hat{p}
\end{align*}
and furthermore
\begin{equation*}
\mc{T} i \mc{T} = -i
\qquad ,\qquad
[\mc{P}, \mc{T}]=0
.\end{equation*}
 For $\mc{PT}$ symmetric Hamiltonians,  eigenvalues are either real or part of a complex conjugate pair, depending on whether the eigenstate in question breaks \PT. The thermodynamics of such systems were studied in \cite{Jones:2009fw}, and this extra symmetry allows us to make physical sense of some (possibly slightly modified) techniques from quantum and classical statistical mechanics.  Of particular interest for us will be that these systems have real partition functions, which is easily seen since it can be expressed completely in terms of real spectral $Z$ functions 
 \begin{equation}
 Z_l = \Tr e^{-l H} = \sum_{i} e^{ - l \lambda_i}  = \frac{1}{2}\sum_{i} \left( e^{ - l \lambda_i}+e^{ - l \lambda^\star_i}\right)
.\end{equation}
However, the semiclassical expression for $Z_l$ given by \eqref{eq:Zl} is no longer a priori true. The integration over the real line is not automatically convergent for $\mc{PT}$ symmetric hamiltonians, and  the integration contour may need to be deformed in the complex plane \cite{Jones:2009fw}. Provided that this can be done, this indeed allows us to consider the thermodynamics of a system in this way without requiring Hermiticity of the Hamiltonian. The particulars of the integration contour in our case will be further discussed below. A wide range of literature on $\mc{PT}$ symmetric quantum mechanics exists, for a pedagogical introduction see for example \cite{Bender:2005tb} and for a more recent review, see \cite{Bender_2015}.

\begin{figure}[h!]
    \centering
        \includegraphics[width=0.6\textwidth]{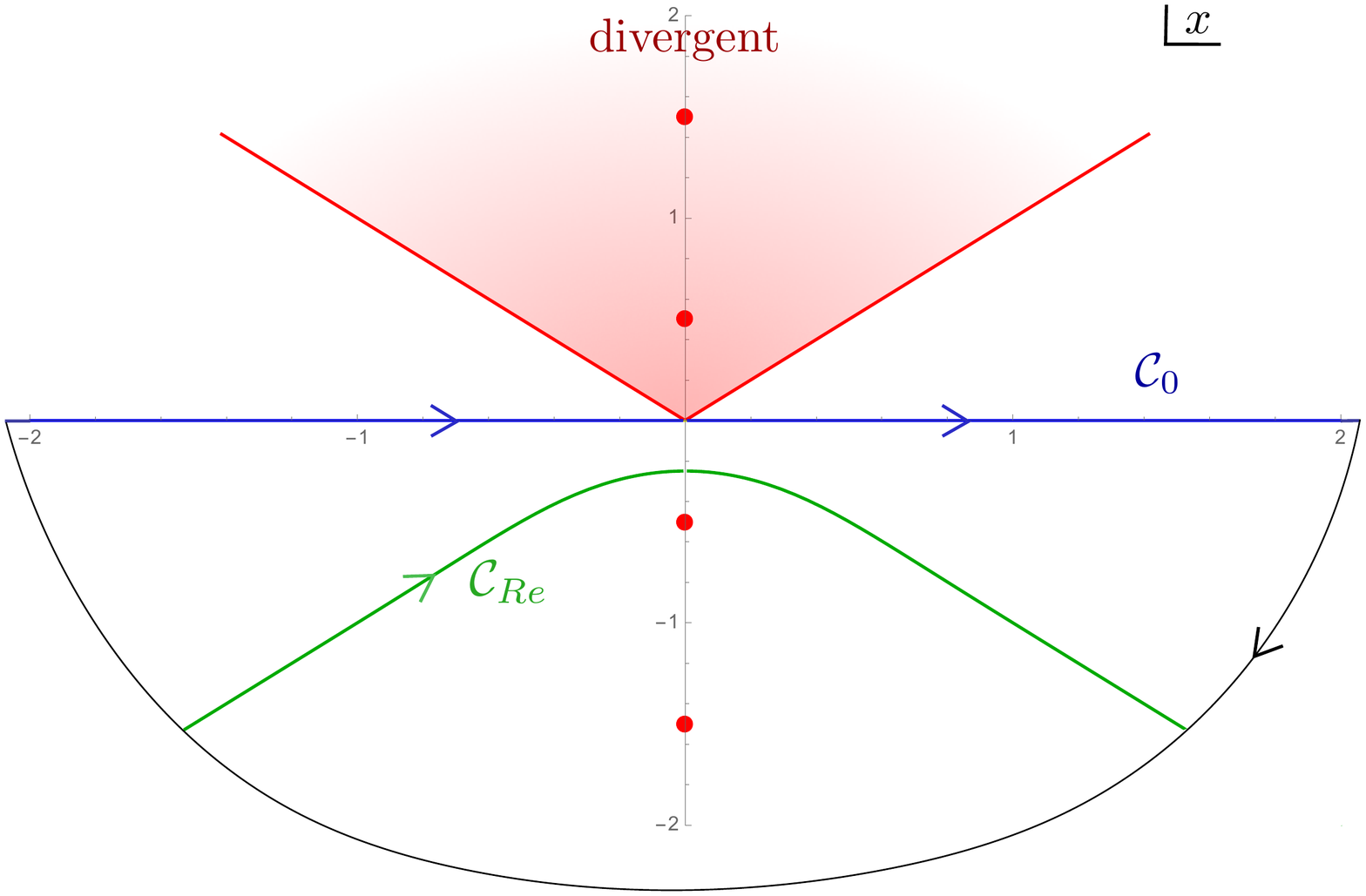}
        \label{fig:Contours}
        \caption{The integration contour in the complex $x$-plane. The blue contour, $\mc{C}_0$ lies along the real axis, whereas the green contour, $\mc{C}_{Re}$ is deformed such that the density matrix, and therefore also the integrand in the expression for the spectral $Z$ functions in \eqref{eq:Zl}, explicitly remains real along the entire contour, here plotted for $ m_2=0.5$ (and $m_1=0$). By closing the contours in the lower half plane, it is clear that these two integration contours gives the same result since the deformation does not cross any poles. The red region corresponds to the wedge where the integrand diverges for large $|x|$. }
    \label{fig:Contours}
\end{figure}

\subsection{Generalised $\mc{PT}$ symmetry  \label{sec:PT_sym_can}}
The density matrix (and Hamiltonian) is invariant under $\mathcal{PT}$ symmetry for certain values of the deformation parameters $  m_1$ and $  m_2$. The easiest such case is $  m_1=0, \,  m_2\neq 0$. In some cases, the density matrix is not invariant under $\mc{PT}$, but rather goes to another density matrix which is related to the original one via a canonical transformation,  
\begin{align*}
\mc{PT}\rho \mc{T}^{-1}\mc{P}^{-1} = \tilde{\rho} \quad \underset{\text{transf.}}{\overset{\text{can.}}{\longrightarrow}} \quad\rho
.\end{align*}
One example is  the case $  m_1=m_2$, where the Hamiltonian is invariant under \PT symmetry followed by the canonical transformation  $x\rightarrow p, \, p\rightarrow -x$. If such a canonical transformation exists, generated by a unitary operator $\mc{U}$, then these systems are invariant under the antiunitary operator $\mc{PTU}$, which is a straightforward generalisation of $\mc{PT}$ symmetry. As long as our canonical transformation raised to some even power gives unity, $\mc{U}^{2k}=1$, the eigenvalues will still split into conjugate pairs \cite{Bender:2002yp} and we can safely compute the spectral $Z$ functions using \eqref{eq:Zl}, provided we can find a suitable integration contour. 


\section{Mass deformations at large $N$ \label{sec:FreeEnergy}}
In this section, we will compute the spectral $Z$ functions for two cases with (generalised) $\mc{PT}$ symmetry, and obtain the large $N$ expression for the free energy of the deformed ABJM theory. The first case corresponds to a only turning on masses for half of the hypermultiplet scalars,  whereas the second case corresponds to either a pure mass or pure FI deformation. These latter situations preserve half of the original supersymmetries. We finish by analysing the case with no apparent \PT ~symmetry.

\subsection{The simplest $\mc{PT}$ symmetric case, $\,  m_1=0$ \label{sec:eta_0}}
We will start by analysing our model in the simplest case where the $\mc{PT}$ symmetry is explicit: $m_1=0$ and $m_2\neq 0$.\footnote{Or, by carrying out the canonical transformation $x\mapsto p, \quad p\mapsto-x$, equivalently $ m_2=0$.}  We then have:
\begin{align*}
U(x)
= 
 \log \left[  2\cosh \pi x  \right]
+ 2\pi i  m_2 x
\qquad,\qquad
T(p)  =
\log\left[ 2
\cosh \pi p
\right]
.\end{align*}

In this case, it is straight-forward, though tedious, to show that there exists a contour in the lower half plane of $x=u+iv\in \mbb{C}$ such that the integrand  of \eqref{eq:Zl} remains real along the entire contour. Explicitly, this contour takes the form of $v=-\frac{i}{\pi}\arctan\left[\frac{\tan(2\pi m_2 |u|)}{\tanh(\pi u)}\right]$ for the appropriate choice of branch cut for the arctan.\footnote{Careful analysis of the branch structure avoids the apparent singularity at $\tan\pi$.} By closing the contour in the lower half plane as illustrated in figure \ref{fig:Contours}, it is clear that in the semiclassical approximation, the Hamiltonian can be recast in a way such that it remains purely real, and as such, $\mc{PT}$ symmetry is unbroken, and all eigenvalues are real.

\subsubsection{The strict thermodynamic limit}
Before we move on to trying to compute the integral in \eqref{eq:Zl}  and $J(\mu)$ explicitly, let us first consider the strict thermodynamical limit of the system, corresponding to the leading order in $N$.
Consider the classical Hamiltonian  in the limit of large $|x|$, $|p|$:
\begin{equation}
H_{cl.}^{N\rightarrow \infty}=   \pi \Big( |p|   + x (\sign\left[\Re (x) \right] +  2 i   m_2  ) \Big)
\end{equation}
Following \cite{Jones:2009fw}, we keep $p$ to be real but analytically continue $x$ to the entire complex plane, i.e. $x=u+iv$. This gives us the Hamiltonian in $p, u,v$-space as:
\begin{align*}
H_{cl.}^{N\rightarrow \infty}
=& 
 \pi \Big( |p|   +|u| - 2   m_2 v+ i \sign\left[u \right] (v +  2    m_2 |u| ) \Big)
\end{align*}
Restricting to the contour where the Hamiltonian is real, $\mc{C}_{Re}$, as discussed under eq \eqref{eq:Zl} gives us an expression for $v$ in terms of $u$ as: 
\begin{equation}
v = - |u| 2  m_2 
\end{equation}
so
\begin{align}
H_{cl.}^{N\rightarrow \infty} 
=&  \pi \Big( |p|   +|u|\left(1 + [2   m_2]^2\right)  \Big)
\quad \leq E
.\end{align}

\begin{figure}
    \centering
        \includegraphics[width=0.5\textwidth]{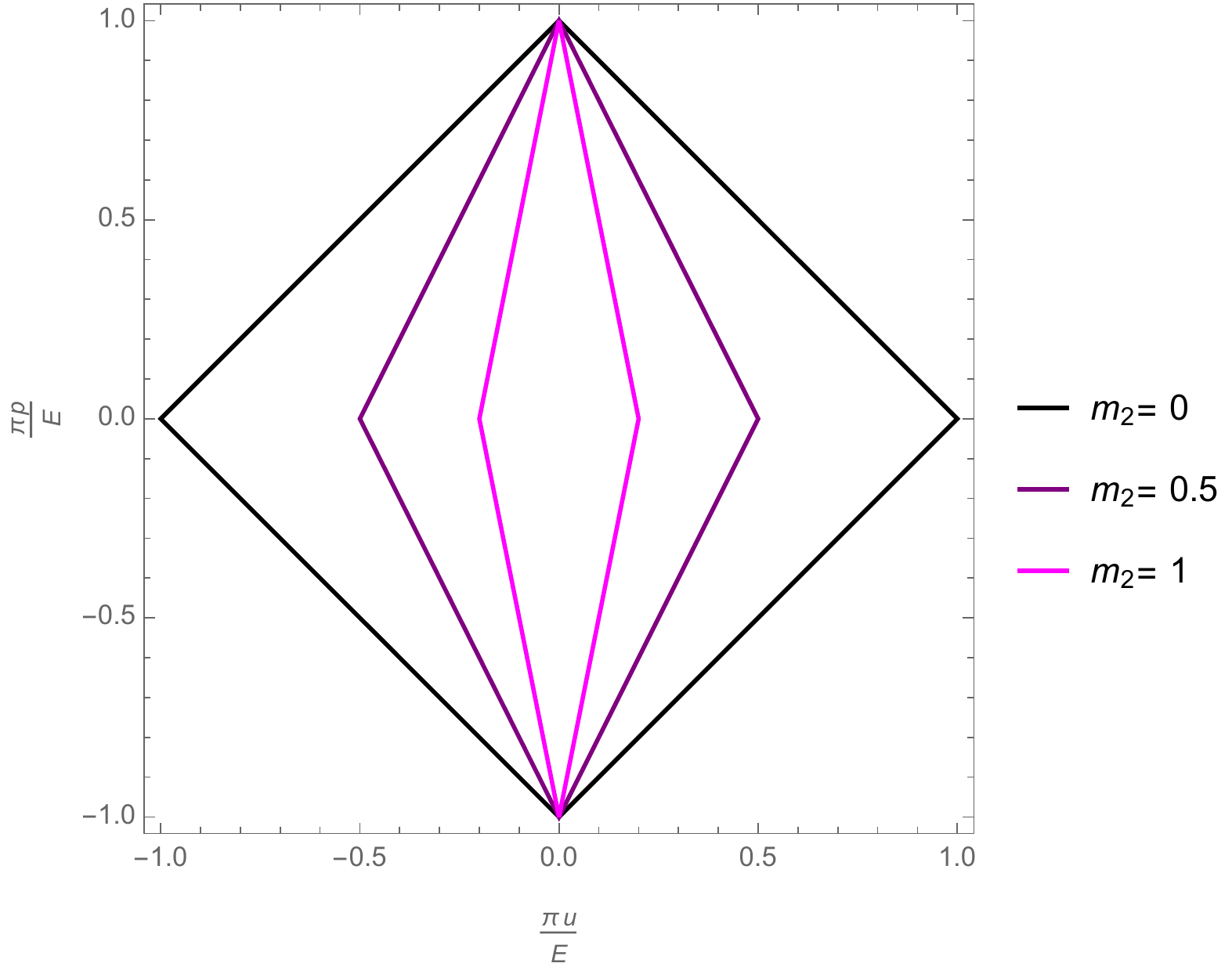}
        \label{fig:Fermik}
    \caption{Fermi surfaces for varying deformation $m_2$ and CS-level $k=1$ plotted as $\pi p/E$, $\pi u/E$. As the deformation increases, the diamond is 'squashed'.}
    \label{fig:FreeEnergy}
\end{figure}

Therefore, the volume of the region of $(u,p)$-space, (which is equivalent to the volume in $(x,p)$-space), with an energy less than or equal to $E$ is given by:
\begin{equation}
\label{eq:vol}
vol(E)=
\frac{2}{\pi^2} \frac{E^2}{\left(1 +\left[2  m_2 \right]^2\right)}
.\end{equation}

By standard thermodynamical arguments, the number of states with energy less than or equal to $E$ is then given by
\begin{equation}
\label{eq:n(E)}
n(E) = \frac{vol(E)}{ 2 \pi \hbar} = \frac{2E^2}{ \pi^2 k \left(1+ \left[2  m_2 \right]^2\right)}.
\end{equation}

The grand canonical potential can equivalently be written in terms of the energy using the density of states $\rho(E)=n'(E)=2CE$  with $C=\frac{2 }{ \pi ^2 k }
\frac{1}{ \left(1+\left[ 2  m_2\right]^2\right)}$:
\begin{align*}
J(\mu)=\int_0^\infty \,dE\,\rho(E) \,\log\left(1+ e^{-E+\mu}\right)
= -2 C\, \text{Li}_3(-e^\mu)
 \overset{\mu\rightarrow\infty}{\approx} \frac{C}{3} \mu^3 
\end{align*}
where $\text{Li}_n(z)$ is a polylogarithm.  This allows us to find the saddle point of the integral representation of the partition function in \eqref{eq:ZinJ} which occurs at:
\begin{align*}
N=\partial_{\mu} J(\mu) = -2 C \, \text{Li}_2(e^{-\mu}) \overset{\mu\rightarrow\infty}{\approx} C \mu^2 
,\end{align*}
which defines $\mu^*(N)=\frac{\sqrt{N}}{\sqrt{C}}$, leading to the free energy as 
\begin{align*}
F=-\log(Z) \overset{\mu\rightarrow\infty}{\approx} - \left(J(\mu^*)-N \mu^* \right)=\frac{2}{3} \frac{N^{3/2}}{\sqrt{C}}.
\end{align*}

Therefore we find a free energy given by
\begin{equation}
\label{eq:Free_E_thermo}
F(N)  \overset{N\rightarrow\infty}{\approx}
\frac{\pi \sqrt{2k}}{3}  N^{3/2} \sqrt{ 1+ \left[2  m_2 \right]^2  }.
\end{equation}

\subsubsection{Beyond  leading order: Airy function behaviour}
There are now two kinds of corrections to our results we should take into account: the first one arises from the classical approximation of the Hamiltonian, and the second from the approximation of this Hamiltonian in the thermodynamic limit. 
This first kind of these corrections will give rise to an expansion of the grand canonical potential in $\hbar$ as:
\begin{align}
\label{eq:J_series}
J(\mu)=\frac{1}{\hbar}J^{(0)}(\mu)+\hbar J^{(1)}(\mu)+\dots
,\end{align}
where the first-order correction can be shown to be independent of the mass parameters (see appendix \ref{sec:quantum}), and we expect higher order corrections to be exponentially suppressed as in the undeformed case. The second type of corrections will  modify $J^{(0)}$ by computing the spectral $Z$ functions exactly rather than the approximate volume of (complex) phase space.  
Let us now consider the corrections arising by exact calculation of the semiclassical spectral $Z$ functions, corresponding to finding subleading corrections in $N$.
These can be obtained in closed form via evaluating the integrals in \eqref{eq:Zl}, i.e.
\begin{equation}
Z_l  \overset{\text{semiclassical}}{=} \; 
\frac{1}{2 \pi \hbar}\int_{-\infty}^{\infty} \hspace*{-2mm}dp \,\frac{1}{\left(2 \cosh(\pi p)\right)^l} \; 
\int_{-\infty}^{\infty}  \hspace*{-2mm} 
dx \, \frac{e^{2\pi i  m_2 x l}}{\left(2\cosh(\pi x)\right)^l} 
,\end{equation}
 using  the relation
  \begin{align*}
 \int_{-\infty}^{\infty}  \hspace*{-2mm} 
dx \, \frac{e^{2\pi i  m_2 x l}}{\left(2\cosh(\pi x)\right)^l}  = 
\frac{\Gamma \left(\frac{1}{2} l (1-2 i  m_2 )\right) \Gamma \left(\frac{1}{2} l (1+2 i  m_2)\right)}{2 \pi  \Gamma (l)}
 .\end{align*}
This gives  the spectral $Z$ functions:
\begin{equation*}
Z_l\approx \frac{1}{\hbar} Z^{(0)}_l 
\end{equation*}
where
\begin{equation}
\label{eq:Zl0}
Z^{(0)}_l = \frac{1}{ (2\pi)^3 } 
\frac{\Gamma \left(\frac{l}{2}\right)^2 \Gamma \left(\frac{l}{2} \left(1-2i  m_2  \right)\right) \Gamma \left(\frac{l}{2}  \left(1+2 i m_2\right)\right)}{\Gamma (l)^2}.
\end{equation}
This will always be real, since $\overline{\Gamma(z)}=\Gamma(\overline{z})$, ensuring  real free energy for our system.

\begin{figure}
    \centering
        \includegraphics[width=0.55\textwidth]{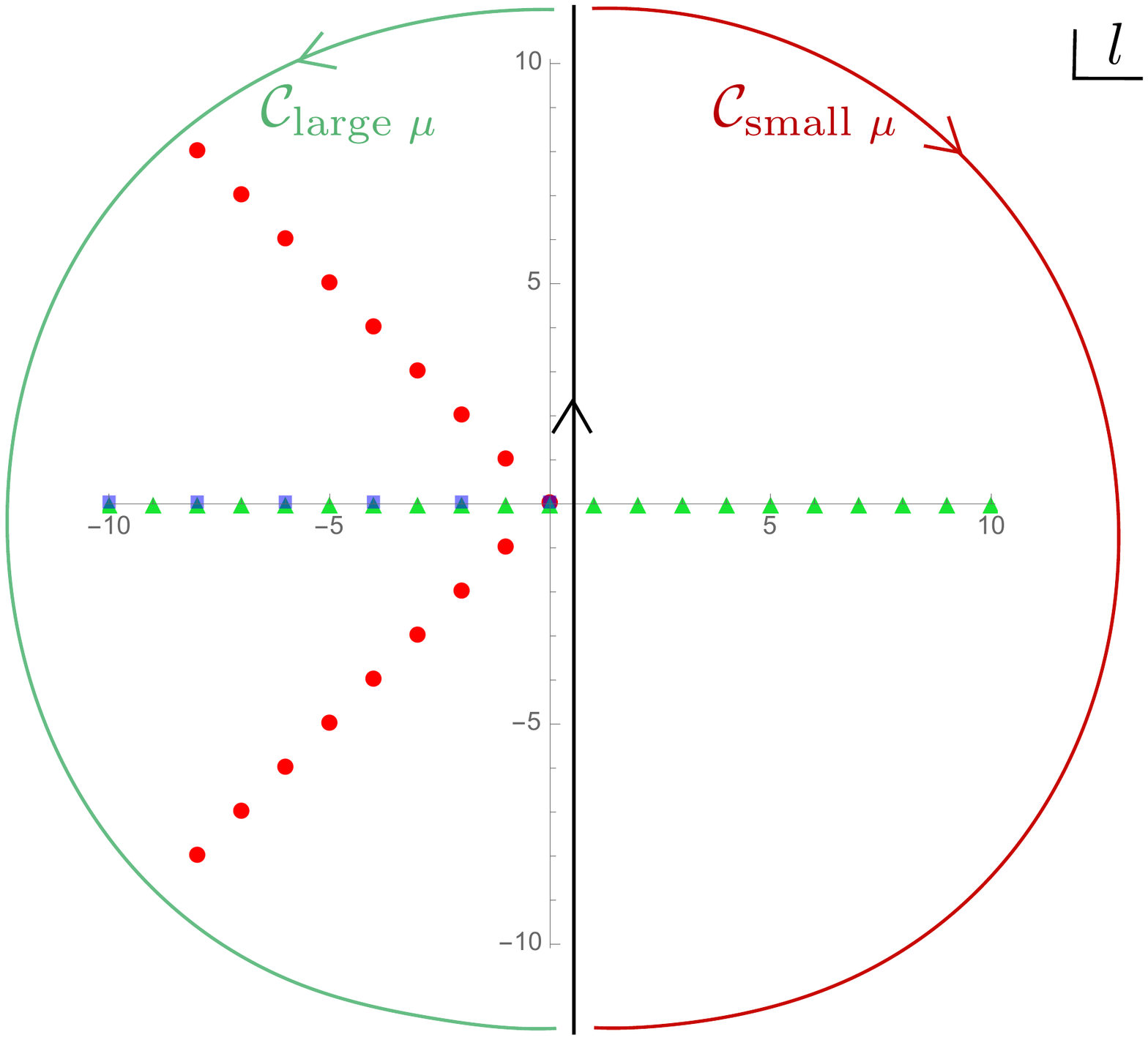}
        \caption{The integration contour in the complex $l$-plane and the poles of the integrand in \eqref{eq:J_MBrep} from the Gamma functions both in \eqref{eq:J_MBrep1} and from $Z_l^{(0)}$ from \eqref{eq:Zl}.  The figure is plotted for $m_1=0,~m_2=1/2$. As $  m_2$ increases, the complex poles from $Z_l^{(0)}$ (red) form a wider angle in the left half-plane. The real poles from $Z_l^{(0)}$ are plotted in blue, whereas the green poles arise from the gamma functions in \eqref{eq:J_MBrep} .}    \label{fig:Contours_MB}
\end{figure}

In the large $N$ limit, the Mellin-Barnes representation of the grand canonical potential \eqref{eq:J_MBrep1} offers us a convenient way to obtain the perturbative part of $J^{(0)}(\mu)$ by only taking into account the residue at the origin,
\begin{equation}
\label{eq:Jpert}
J^{(0)}_{pert}(\mu)=
\frac{2  \mu ^3}{3 \pi ^2 k }
\frac{1}{ \left(1+\left[ 2  m_2\right]^2\right)}
+
\frac{\mu }{3  k}
\frac{   \left(1-\frac{1}{2}\left[ 2  m_2\right]^2\right)
}{
  \left(1+\left[ 2  m_2\right]^2\right)
}
+
\frac{ 2  \zeta (3)}{k\pi^2}
\frac{
 \left(1+ \frac{1}{2}\left[ 2  m_2\right]^2\right)
 }{
 \left(1+\left[ 2  m_2\right]^2\right)
 }.
\end{equation}
This allows us to compute the partition function  via equation \eqref{eq:ZinJ}, where the integral is evaluated exactly using the integral representation of the Airy function,
\begin{equation*}
\frac{1}{2\pi i}\int_{\mc{C}} d\mu\, e^{\frac{C}{3}\mu^3+(B-N) \mu +A} = e^{A}C^{-1/3}\text{Ai} \left[ C^{-1/3} (N-B)\right]
,\end{equation*}
giving us:
\begin{align*}
Z_{pert.}(N)= e^{A}C^{-1/3}\text{Ai} \left[ C^{-1/3} (N-B)\right]
\end{align*}
where
\begin{align*}
C=\frac{2 }{ \pi ^2 k }
\frac{1}{ \left(1+\left[ 2  m_2\right]^2\right)}
\quad,\quad
B= \frac{1}{3  k}
\frac{   \left(1-\frac{1}{2}\left[ 2  m_2\right]^2\right)
}{
  \left(1+\left[ 2  m_2\right]^2\right)
}
\quad,\quad
A=
\frac{ 2\zeta (3)}{k\pi^2}
\frac{
 \left(1+ \frac{1}{2}\left[ 2  m_2\right]^2\right)
 }{
 \left(1+\left[ 2  m_2\right]^2\right)
 }
.\end{align*}

This gives the free energy as
\small{
\begin{align}
F=-\log(Z_{pert}(N))\overset{N\rightarrow \infty}{\rightarrow}\,
  \frac{\pi \sqrt{2 k}}{3}\, N^{3/2}\,  \sqrt{1+\left[ 2  m_2\right]^2}
-
\sqrt{N} \frac{\pi  }{3 \sqrt{2 k} }
\frac{ \left(1-\frac{1}{2 } \left[ 2  m_2\right]^2 \right)
}{
 \sqrt{1+\left[ 2  m_2\right]^2}
 }
+\mc{O}(\log N)
\end{align}
}
where the first term precisely reproduces the result given by the polygonal approximation in equation \eqref{eq:Free_E_thermo}.

\subsection{The maximally supersymmetric deformation: $m_1=m_2=m$} 
As discussed in section \ref{sec:PT_sym_can}, the Hamiltonian has generalised $\mc{PT}$ symmetry for $ m_1=  m_2=m$. This represents the case where all hypermultiplet scalars are given the same mass, and preserves $\mc{N}=6$ supersymmetry, breaking only the conformal part of the algebra.\footnote{Another, equally symmetric, situation would be the case of pure FI deformation, where $m_1=-m_2=2\zeta$.} Recall by equation \eqref{eq:Hamiltonian_def}, we have:

\begin{equation*}
\rho = e^{-\frac{1}{2}U(x)} e^{-T(p)} e^{-\frac{1}{2}U(x)} 
   \;  =e^{-\hat{H}}
\end{equation*}
where now $U(x),\, T(p)$: 
\begin{align*}
U(x)
= 
 \log \left[ 2 \cosh \pi x  \right]
+2 \pi i \,  m x
\qquad,\qquad
T(p)  =
\log\left[ 2
\cosh \pi p
\right]
+ 2 \pi i \,  m p
.\end{align*}

In this case, the classical Hamiltonian  is symmetric in $x,p$ and given by:
\begin{align*}
H_{cl.} = \log 2 \cosh  \pi x +\log 2 \cosh \pi p + 2 \pi  i m (x+p)
.\end{align*}
Repeating the calculation of section \ref{sec:eta_0} gives 
\begin{equation*}
Z^{(0)}_l = \frac{1}{(2\pi)^3 } \,
\frac{ \Gamma^2 \left(\frac{l}{2} \left(1- 2 i  m \right)\right) \Gamma^2 \left(\frac{l}{2}  \left(1+2 i  m\right)\right)}{\Gamma (l)^2}
\end{equation*}
where we again can use the Mellin-Barnes representation to obtain the perturbative part of the partition function as
\begin{align*}
Z_{pert.}(N)= e^{A}C^{-1/3}\text{Ai} \left[ C^{-1/3} (N-B)\right]
\end{align*}
but now with the parameters
\begin{align*}
C=\frac{2}{\pi ^2 k }\frac{1}{\left(1+\left[ 2  m \right] ^2\right)^2}
\quad,\quad
B=\frac{1}{3k}
\frac{\left(1-\left[ 2  m \right] ^2\right)}{ \left(1+\left[ 2  m \right] ^2\right)^2}
\quad,\quad
A=
\frac{2 \zeta (3)}{\pi ^2 k}\frac{1}{ \left(\left[ 2  m \right]^2+1\right)}
.\end{align*}

This gives the free energy as
\small{
\begin{align}
F=-\log(Z_{pert}(N))\overset{N\rightarrow \infty}{\rightarrow}\,
\frac{ \pi \sqrt{2k}}{3} N^{3/2} \left(1+\left[ 2  m \right]^2\right) 
-  \frac{\pi }{3\sqrt{2k}} \sqrt{N}\,
\frac{  \left(1-\left[ 2  m \right]^2\right) }{ \left(1+\left[ 2  m \right]^2\right)} 
+\mc{O}(\log N)
\end{align}
}
where $\sqrt{N}$-term will receive corrections from the next-to-leading order in the WKB approximation, which are independent of $m$, and higher-order corrections will be of at least order one.

\subsection{Beyond $\mc{PT}$ symmetry}
The integrals for computing the semiclassical spectral $Z$ functions in \eqref{eq:Zl} actually converge for all values of $  m_1,\,  m_2$. However, the physical interpretation of this situation is more subtle. We obtain spectral $Z$ functions
\begin{equation}
\label{eq:genZ0}
Z^{(0)}_l = \frac{1}{(2\pi)^3 } \,
\frac{ \Gamma \left(\frac{l}{2} \left(1- 2 i  m_1 \right)\right) \Gamma \left(\frac{l}{2}  \left(1+2 i  m_1\right)\right) 
 \Gamma \left(\frac{l}{2} \left(1- 2 i  m_2 \right)\right) \Gamma \left(\frac{l}{2}  \left(1+2 i  m_2\right)\right)
 }{\Gamma (l)^2}
\end{equation}
leading to an Airy function behaviour of the partition function, but now the parameters $A,B$ and $C$ as expected depend on both $  m_2$ and $  m_1$ via:
{\footnotesize
\begin{align*}
& \qquad\qquad\qquad\qquad\qquad
C=\frac{2 }{\pi ^2 k }
\frac{1}{ \left(1+ \left[2 m_1\right] ^2\right)  \left(1+ \left[2 m_2 \right]^2\right)}
\\
B&=\frac{1 }{3k}\frac{ \left(1-\frac{1}{2}\left[ \left[2 m_1\right] ^2+ \left[2 m_2\right] ^2\right]\right)}{ \left(1+ \left[2 m_1\right] ^2\right)  \left(1+ \left[2 m_2\right] ^2\right)}
\qquad,\qquad
A=
\frac{2\zeta (3)}{\pi ^2 k}
\frac{ \left(1+\frac{1}{2}\left[ \left[2 m_1\right] ^2+ \left[2 m_2\right] ^2\right]\right)}{ \left(1+ \left[2 m_1\right] ^2\right) \left(1+ \left[2 m_2\right] ^2\right)}
.\end{align*}
}
For the partition function, we as before get
\begin{align*}
Z_{pert}( m_2,  m_1) = e^A C^{-\frac{1}{3} } \text{Ai} \Big[ C^{-\frac{1}{3} }  (N-B)\Big]
\end{align*}
and the free energy in the large $N$ limit as
\begin{align*}
F_{pert}=\frac{2 N^{3/2}}{3 \sqrt{C }}-\frac{B \sqrt{N}}{\sqrt{C}}+ \frac{1}{4}\log \left( 16 \pi^2  C N \right)-A+\mc{O}\left(\frac{1}{\sqrt{N}}\right).
\end{align*}
The free energy in terms of the deformation parameters is 
\begin{align}
\label{eq:F_loc_general}
F_{pert}=&\frac{\pi  \,\sqrt{2 k}   }{3}N^{3/2}
\sqrt{\left(1+ \left[2 m_1\right] ^2\right)  \left(1+ \left[2 m_2 \right]^2\right)}\nonumber \\
&-
\frac{\pi }{3\sqrt{2 k}}\sqrt{N}\frac{1-\frac{1}{2}\left[ \left[2 m_1\right] ^2+ \left[2 m_2\right] ^2\right]}{\sqrt{\left(1+ \left[2 m_1\right] ^2\right)  \left(1+ \left[2 m_2\right] ^2\right)}}
+\mathcal{O}(\log N).
\end{align}
The leading term agrees with the results of \cite{Freedman:2013ryh,Nosaka:2015iiw}, after a straightforward analytic continuation.

\section{Discussion \label{sec:Discussion}}

In this paper we have computed the free energy of real mass deformed ABJM theory without analytic continuation using  generalised \PT~symmetry. The final results are compatible with previous results from $F$ maximisation under the simple assumption that the partition function is analytic in deformation parameters. An interesting aspect is that the partition function remains real even when generalised \PT~symmetry is not present at the level of the free fermion Hamiltonian.

It would be interesting to study nonperturbative corrections further in light of  \cite{Honda:2018cky}, which claims that for some values of the deformation parameters these are no longer exponentially suppressed in $N$. It would also be interesting to study the conjectured supersymmetry breaking from the bulk perspective, however, our analysis shows no sign of it.

 In the case that has  \PT~symmetry, $m_1=0$, we explicitly show in appendix \ref{sec:quantum} that the first quantum correction to the Hamiltonian is also \PT~invariant. Noting that the full quantum Hamiltonian only contains even powers of $\hbar$, one can show that for general $T(p)$ and $U(x)$ such that $T(p)$ is real and  $U(x)=f_{even}(x)+if_{odd}(x)$,  this symmetry persists to all orders in $\hbar$. This guarantees that the full quantum Hamiltonian is \PT~symmetric when $m_1=0$.
 
We also demonstrate in appendix \ref{sec:quantum} that the first quantum correction to the spectral $Z$ function is real for all values of the deformation parameters, even though there is no obvious generalised \PT~symmetry.  It would also be extremely enlightening to have a physical explanation of the reality of the partition function in this case.

An obvious extension of our work would be to investigate deformations of more general quiver theories and see whether generalised \PT~symmetry plays a role there as well. It would also be interesting to consider results at finite $N$. This is particularly interesting considering the second term in \eqref{eq:F_loc_general} changes sign at $[2 m_1] ^2+ [2 m_2] ^2=2$, where \cite{Honda:2018cky} conjectures breaking of supersymmetry.

\section*{Acknowledgments}
It is a pleasure to thank Carl M.~Bender and Christopher\hspace*{-2mm} {\greektext{ Triantafull\'akhs}} Rosen for insightful discussions. LA is supported  by the  EPSRC programme grant ``New Geometric Structures from String Theory'', EP/K034456/1. The work of MMR is supported by the European Research Council under the European Union's Seventh Framework Programme (FP7/2007-2013), ERC Grant agreement ADG 339140. LA is grateful for the hospitality of the Galileo Galilei Institute during the workshop ``Supersymmetric Quantum Field Theories in the Non-perturbative Regime''.

\appendix

\section{Quantum corrections \label{sec:quantum}}
The leading quantum corrections to the free energy will be independent of the deformation parameters, which we can see by explicit calculation of the first-order quantum corrections to the spectral $Z$ functions in \eqref{eq:Zl_series}.  In \cite{Marino:2011eh}, it was shown that for a general Hamiltonian defined via \eqref{eq:Hw}, the quantum corrections can be obtained in a systematic way using a generalisation \cite{Voros:1976sm,  Grammaticos:1978tf} of the standard Wigner-Kirkwood expansion \cite{Wigner:1932aa, Kirkwood:1933aa}. The first order quantum to the spectral $Z$ functions is given by:
{\small
\begin{align}
\label{eq:Zl1}
Z^{(1)}_l =&\,
\frac{l}{2 \pi }\int_{-\infty}^{\infty} \hspace*{-2mm}dp \,\int_{-\infty}^{\infty}  
\hspace*{-2mm} dx \, e^{-l H_{cl.} (x,p)}
\left(
\frac{1}{24} \,U'(x)^2T''(p)-\frac{1}{12} \,T'(p)^2U''(x)
\right)
\\
\nonumber
&+\,
\frac{1}{2 \pi }\int_{-\infty}^{\infty} \hspace*{-2mm}dp \,\int_{-\infty}^{\infty}  
\hspace*{-2mm} dx \, e^{-l H_{cl.} (x,p)}
\left[
\frac{l^3}{24}\left(
 \,U'(x)^2T''(p)+U''(x) T'(p)^2\,
\right)
-\frac{l^2}{8}\, U''(x)T''(p)
\right]
.\end{align}
}
This can be computed for our general Hamiltonian using the following integrals:
\begin{align*}
\frac{\pi^2}{2^l}\int_{-\infty}^{\infty}\hspace*{-2mm} dz\,
 \frac{e^{-2\pi i m\, l z}}{\left(\cosh{\pi z}\right)^l}\left(2im+\tanh(\pi z)\right)^2
=&\,
\frac{2\pi}{ l} \,\frac{\Gamma\left(\frac{l}{2}(1+2im) +1 \right) \, \Gamma\left(\frac{l}{2}(1-2im) +1 \right)}{\Gamma(l+2)}
\\
\frac{\pi^2}{2^l}\int_{-\infty}^{\infty}
\hspace*{-2mm} dz\,
 \frac{e^{-2\pi i m \, l z}}{\left(\cosh{\pi z}\right)^{l+2}}\
=&\,
2\pi\,
\frac{
\Gamma\left(\frac{l}{2}(1+2im) +1 \right) \, \Gamma\left(\frac{l}{2}(1-2im) +1 \right)
 }{  \Gamma (l+2)}
,\end{align*}
together with 
\begin{align*}
T'(p)^2
=
\pi^2 \left(2i m_2+\tanh(\pi x)\right)
\qquad,&\qquad
U'(x)^2
=
\pi^2 \left(2i m_1+\tanh(\pi p)\right)
\\
T''(p)
=
\frac{\pi^2}{\cosh^2(\pi p)}
\qquad,&\qquad
U''(x)
=
\frac{\pi^2}{\cosh^2(\pi x)}
.\end{align*}

Combining all of this, we use \eqref{eq:Zl1} to find 
\begin{align*}
Z^{(1)}_l =&\,
-\frac{\pi ^4 (l-1) l^2 \left(1+\left[2 m_1\right]^2\right) \left(1+\left[2 m_2\right]^2\right)}{24 (l+1)}
Z^{(0)}_l
\end{align*}
where $Z^{(0)}_l$ is given by \eqref{eq:genZ0}. Using the Mellin-Barnes representation to compute the grand canonical potential, \eqref{eq:J_MBrep}, in the large $N$ limit gives a correction to the grand canonical potential via \eqref{eq:J_series}, giving  $J^{(1)}(\mu)$ as:
\begin{align*}
\hbar J^{(1)}(\mu)=- \hbar 
\, \, \text{Res}_{l=0} \left\{\, \Gamma(l) \Gamma(-l) Z^{(1)}_l \, e^{l \mu}\right\}=
 \mu\frac{k }{24} - \frac{k}{12} 
,\end{align*}
precisely reproducing the result of \cite{Marino:2011eh}, with no dependence on the deformation parameters.

%

\addcontentsline{toc}{section}{Bibliography}
\bibliographystyle{JHEP}
\bibliography{HoloSF}

\providecommand{\href}[2]{#2}\begingroup\raggedright\begin{thebibliography}{10}

\bibitem{Aharony:2008ug}
O.~Aharony, O.~Bergman, D.~L. Jafferis, and J.~Maldacena, {\it {N=6
  superconformal Chern-Simons-matter theories, M2-branes and their gravity
  duals}},  {\em JHEP} {\bf 0810} (2008) 091,
  [\href{http://xxx.lanl.gov/abs/0806.1218}{{\tt arXiv:0806.1218}}].

\bibitem{Klebanov:1996un}
I.~R. Klebanov and A.~A. Tseytlin, {\it {Entropy of near extremal black
  p-branes}},  {\em Nucl. Phys.} {\bf B475} (1996) 164--178,
  [\href{http://xxx.lanl.gov/abs/hep-th/9604089}{{\tt hep-th/9604089}}].

\bibitem{Drukker:2010nc}
N.~Drukker, M.~Marino, and P.~Putrov, {\it {From weak to strong coupling in
  ABJM theory}},  {\em Commun. Math. Phys.} {\bf 306} (2011) 511--563,
  [\href{http://xxx.lanl.gov/abs/1007.3837}{{\tt arXiv:1007.3837}}].

\bibitem{Hosomichi:2008jd}
K.~Hosomichi, K.-M. Lee, S.~Lee, S.~Lee, and J.~Park, {\it {N=4 Superconformal
  Chern-Simons Theories with Hyper and Twisted Hyper Multiplets}},  {\em JHEP}
  {\bf 07} (2008) 091, [\href{http://xxx.lanl.gov/abs/0805.3662}{{\tt
  arXiv:0805.3662}}].

\bibitem{Hosomichi:2008jb}
K.~Hosomichi, K.-M. Lee, S.~Lee, S.~Lee, and J.~Park, {\it {N=5,6
  Superconformal Chern-Simons Theories and M2-branes on Orbifolds}},  {\em
  JHEP} {\bf 09} (2008) 002, [\href{http://xxx.lanl.gov/abs/0806.4977}{{\tt
  arXiv:0806.4977}}].

\bibitem{Gomis:2008vc}
J.~Gomis, D.~Rodriguez-Gomez, M.~Van~Raamsdonk, and H.~Verlinde, {\it {A
  Massive Study of M2-brane Proposals}},  {\em JHEP} {\bf 09} (2008) 113,
  [\href{http://xxx.lanl.gov/abs/0807.1074}{{\tt arXiv:0807.1074}}].

\bibitem{Pestun:2007rz}
V.~Pestun, {\it {Localization of gauge theory on a four-sphere and
  supersymmetric Wilson loops}},  {\em Commun. Math. Phys.} {\bf 313} (2012)
  71--129, [\href{http://xxx.lanl.gov/abs/0712.2824}{{\tt arXiv:0712.2824}}].

\bibitem{Kapustin:2009kz}
A.~Kapustin, B.~Willett, and I.~Yaakov, {\it {Exact Results for Wilson Loops in
  Superconformal Chern-Simons Theories with Matter}},  {\em JHEP} {\bf 03}
  (2010) 089, [\href{http://xxx.lanl.gov/abs/0909.4559}{{\tt
  arXiv:0909.4559}}].

\bibitem{Jafferis:2010un}
D.~L. Jafferis, {\it {The Exact Superconformal R-Symmetry Extremizes Z}},  {\em
  JHEP} {\bf 05} (2012) 159, [\href{http://xxx.lanl.gov/abs/1012.3210}{{\tt
  arXiv:1012.3210}}].

\bibitem{Hama:2010av}
N.~Hama, K.~Hosomichi, and S.~Lee, {\it {Notes on SUSY Gauge Theories on
  Three-Sphere}},  {\em JHEP} {\bf 03} (2011) 127,
  [\href{http://xxx.lanl.gov/abs/1012.3512}{{\tt arXiv:1012.3512}}].

\bibitem{Pestun:2016zxk}
V.~Pestun et~al., {\it {Localization techniques in quantum field theories}},
  {\em J. Phys.} {\bf A50} (2017), no.~44 440301,
  [\href{http://xxx.lanl.gov/abs/1608.0295}{{\tt arXiv:1608.0295}}].

\bibitem{Kapustin:2010xq}
A.~Kapustin, B.~Willett, and I.~Yaakov, {\it {Nonperturbative Tests of
  Three-Dimensional Dualities}},  {\em JHEP} {\bf 10} (2010) 013,
  [\href{http://xxx.lanl.gov/abs/1003.5694}{{\tt arXiv:1003.5694}}].

\bibitem{Marino:2011eh}
M.~Marino and P.~Putrov, {\it {ABJM theory as a Fermi gas}},  {\em J. Stat.
  Mech.} {\bf 1203} (2012) P03001,
  [\href{http://xxx.lanl.gov/abs/1110.4066}{{\tt arXiv:1110.4066}}].

\bibitem{Bender:1998ke}
C.~M. Bender and S.~Boettcher, {\it {Real spectra in nonHermitian Hamiltonians
  having PT symmetry}},  {\em Phys. Rev. Lett.} {\bf 80} (1998) 5243--5246,
  [\href{http://xxx.lanl.gov/abs/physics/9712001}{{\tt physics/9712001}}].

\bibitem{Bender:2002yp}
C.~M. Bender, M.~V. Berry, and A.~Mandilara, {\it {Generalized PT symmetry and
  real spectra}},  {\em J. Phys.} {\bf A35} (2002) L467.

\bibitem{Drukker:2015awa}
N.~Drukker and J.~Felix, {\it {3d mirror symmetry as a canonical
  transformation}},  {\em JHEP} {\bf 05} (2015) 004,
  [\href{http://xxx.lanl.gov/abs/1501.0226}{{\tt arXiv:1501.0226}}].

\bibitem{Marino:2016new}
M.~Marino, {\it {Localization at large N in Chern--Simons-matter theories}},
  {\em J. Phys.} {\bf A50} (2017), no.~44 443007,
  [\href{http://xxx.lanl.gov/abs/1608.0295}{{\tt arXiv:1608.0295}}].

\bibitem{Freedman:2013ryh}
D.~Z. Freedman and S.~S. Pufu, {\it {The holography of $F$-maximization}},
  {\em JHEP} {\bf 03} (2014) 135,
  [\href{http://xxx.lanl.gov/abs/1302.7310}{{\tt arXiv:1302.7310}}].

\bibitem{Nosaka:2015iiw}
T.~Nosaka, {\it {Instanton effects in ABJM theory with general R-charge
  assignments}},  {\em JHEP} {\bf 03} (2016) 059,
  [\href{http://xxx.lanl.gov/abs/1512.0286}{{\tt arXiv:1512.0286}}].

\bibitem{Honda:2018cky}
M.~Honda, T.~Nosaka, K.~Shimizu, and S.~Terashima, {\it {Supersymmetry Breaking
  in a Large N Gauge Theory with Gravity Dual}},
  \href{http://xxx.lanl.gov/abs/1807.0887}{{\tt arXiv:1807.0887}}.

\bibitem{Benna:2009xd}
M.~K. Benna, I.~R. Klebanov, and T.~Klose, {\it {Charges of Monopole Operators
  in Chern-Simons Yang-Mills Theory}},  {\em JHEP} {\bf 01} (2010) 110,
  [\href{http://xxx.lanl.gov/abs/0906.3008}{{\tt arXiv:0906.3008}}].

\bibitem{Closset:2012vg}
C.~Closset, T.~T. Dumitrescu, G.~Festuccia, Z.~Komargodski, and N.~Seiberg,
  {\it {Contact Terms, Unitarity, and F-Maximization in Three-Dimensional
  Superconformal Theories}},  {\em JHEP} {\bf 10} (2012) 053,
  [\href{http://xxx.lanl.gov/abs/1205.4142}{{\tt arXiv:1205.4142}}].

\bibitem{Gulotta:2012yd}
D.~R. Gulotta, C.~P. Herzog, and T.~Nishioka, {\it {The ABCDEF's of Matrix
  Models for Supersymmetric Chern-Simons Theories}},  {\em JHEP} {\bf 04}
  (2012) 138, [\href{http://xxx.lanl.gov/abs/1201.6360}{{\tt
  arXiv:1201.6360}}].

\bibitem{Anderson:2014hxa}
L.~Anderson and K.~Zarembo, {\it {Quantum Phase Transitions in Mass-Deformed
  ABJM Matrix Model}},  {\em JHEP} {\bf 09} (2014) 021,
  [\href{http://xxx.lanl.gov/abs/1406.3366}{{\tt arXiv:1406.3366}}].

\bibitem{Anderson:2015ioa}
L.~Anderson and J.~G. Russo, {\it {ABJM Theory with mass and FI deformations
  and Quantum Phase Transitions}},  {\em JHEP} {\bf 05} (2015) 064,
  [\href{http://xxx.lanl.gov/abs/1502.0682}{{\tt arXiv:1502.0682}}].

\bibitem{Nosaka:2015bhf}
T.~Nosaka, K.~Shimizu, and S.~Terashima, {\it {Large N behavior of mass
  deformed ABJM theory}},  {\em JHEP} {\bf 03} (2016) 063,
  [\href{http://xxx.lanl.gov/abs/1512.0024}{{\tt arXiv:1512.0024}}].

\bibitem{Nosaka:2016vqf}
T.~Nosaka, K.~Shimizu, and S.~Terashima, {\it {Mass Deformed ABJM Theory on
  Three Sphere in Large N limit}},  {\em JHEP} {\bf 03} (2017) 121,
  [\href{http://xxx.lanl.gov/abs/1608.0265}{{\tt arXiv:1608.0265}}].

\bibitem{Wigner:1932aa}
E.~Wigner, {\it On the quantum correction for thermodynamic equilibrium},  {\em
  Physical Review} {\bf 40} (1932), no.~5 749--759.

\bibitem{1984PhR...106..121H}
M.~{Hillery}, R.~F. {O'Connell}, M.~O. {Scully}, and E.~P. {Wigner}, {\it
  {Distribution functions in physics: Fundamentals}},  {\em Physics Reports}
  {\bf 106} (Apr., 1984) 121--167.

\bibitem{Hatsuda:2015oaa}
Y.~Hatsuda, {\it {Spectral zeta function and non-perturbative effects in ABJM
  Fermi-gas}},  {\em JHEP} {\bf 11} (2015) 086,
  [\href{http://xxx.lanl.gov/abs/1503.0788}{{\tt arXiv:1503.0788}}].

\bibitem{Hatsuda:2012hm}
Y.~Hatsuda, S.~Moriyama, and K.~Okuyama, {\it {Exact Results on the ABJM Fermi
  Gas}},  {\em JHEP} {\bf 10} (2012) 020,
  [\href{http://xxx.lanl.gov/abs/1207.4283}{{\tt arXiv:1207.4283}}].

\bibitem{Putrov:2012zi}
P.~Putrov and M.~Yamazaki, {\it {Exact ABJM Partition Function from TBA}},
  {\em Mod. Phys. Lett.} {\bf A27} (2012) 1250200,
  [\href{http://xxx.lanl.gov/abs/1207.5066}{{\tt arXiv:1207.5066}}].

\bibitem{Hatsuda:2012dt}
Y.~Hatsuda, S.~Moriyama, and K.~Okuyama, {\it {Instanton Effects in ABJM Theory
  from Fermi Gas Approach}},  {\em JHEP} {\bf 01} (2013) 158,
  [\href{http://xxx.lanl.gov/abs/1211.1251}{{\tt arXiv:1211.1251}}].

\bibitem{Calvo:2012du}
F.~Calvo and M.~Marino, {\it {Membrane instantons from a semiclassical TBA}},
  {\em JHEP} {\bf 05} (2013) 006,
  [\href{http://xxx.lanl.gov/abs/1212.5118}{{\tt arXiv:1212.5118}}].

\bibitem{Hatsuda:2013gj}
Y.~Hatsuda, S.~Moriyama, and K.~Okuyama, {\it {Instanton Bound States in ABJM
  Theory}},  {\em JHEP} {\bf 05} (2013) 054,
  [\href{http://xxx.lanl.gov/abs/1301.5184}{{\tt arXiv:1301.5184}}].

\bibitem{Hatsuda:2013oxa}
Y.~Hatsuda, M.~Marino, S.~Moriyama, and K.~Okuyama, {\it {Non-perturbative
  effects and the refined topological string}},  {\em JHEP} {\bf 09} (2014)
  168, [\href{http://xxx.lanl.gov/abs/1306.1734}{{\tt arXiv:1306.1734}}].

\bibitem{Jones:2009fw}
H.~F. Jones and E.~S. Moreira, Jr, {\it {Quantum and Classical Statistical
  Mechanics of a Class of non-Hermitian Hamiltonians}},  {\em J. Phys.} {\bf
  A43} (2010) 055307, [\href{http://xxx.lanl.gov/abs/0905.2879}{{\tt
  arXiv:0905.2879}}].

\bibitem{Bender:2005tb}
C.~M. Bender, {\it {Introduction to PT-Symmetric Quantum Theory}},  {\em
  Contemp. Phys.} {\bf 46} (2005) 277--292,
  [\href{http://xxx.lanl.gov/abs/quant-ph/0501052}{{\tt quant-ph/0501052}}].

\bibitem{Bender_2015}
C.~M. Bender, {\it Pt-symmetric quantum theory},  {\em Journal of Physics:
  Conference Series} {\bf 631} (Jul, 2015) 012002.

\bibitem{Voros:1976sm}
A.~Voros, {\it {Asymptotic $\hbar$-Expansions of Stationary Quantum States}},
  {\em Ann. Inst. H. Poincare Phys. Theor.} {\bf 26} (1977) 343--403.

\bibitem{Grammaticos:1978tf}
B.~Grammaticos and A.~Voros, {\it {Semiclassical Approximations for Nuclear
  Hamiltonians. 1. Spin Independent Potentials}},  {\em Annals Phys.} {\bf 123}
  (1979) 359.

\bibitem{Kirkwood:1933aa}
J.~G. Kirkwood, {\it Quantum statistics of almost classical assemblies},  {\em
  Physical Review} {\bf 44} (1933), no.~1 31--37.

\end{thebibliography}\endgroup

\end{document}